\documentclass[aps,prl,twocolumn,floats,showpacs,floatfix,a4paper]{revtex4}
\usepackage{epsfig}
\usepackage{color}
\usepackage{bm}
\usepackage{latexsym}

\begin{document}
\newcommand{\cc}{{\bf\Large C }}
\newcommand{\hide}[1]{}
\newcommand{\tbox}[1]{\mbox{\tiny #1}}
\newcommand{\half}{\mbox{\small $\frac{1}{2}$}}
\newcommand{\sinc}{\mbox{sinc}}
\newcommand{\const}{\mbox{const}}
\newcommand{\trc}{\mbox{trace}}
\newcommand{\intt}{\int\!\!\!\!\int }
\newcommand{\ointt}{\int\!\!\!\!\int\!\!\!\!\!\circ\ }
\newcommand{\eexp}{\mbox{e}^}
\newcommand{\EPS} {\mbox{\LARGE $\epsilon$}}
\newcommand{\ar}{\mathsf r}
\newcommand{\im}{\mbox{Im}}
\newcommand{\re}{\mbox{Re}}
\newcommand{\bmsf}[1]{\bm{\mathsf{#1}}}
\newcommand{\dd}[1]{\:\mbox{d}#1}
\newcommand{\abs}[1]{\left|#1\right|}
\newcommand{\bra}[1]{\left\langle #1\right|}
\newcommand{\ket}[1]{\left|#1\right\rangle }
\newcommand{\mbf}[1]{{\mathbf #1}}
\newcommand{\eos}{\,.}
\definecolor{red}{rgb}{1,0.0,0.0}

\title{Short-Time Loschmidt Gap in Dynamical Systems with Critical Chaos}
\author{Carl T. West$^{1,3}$, Tomaz Prosen$^2$ and Tsampikos Kottos$^{1,3}$}

\affiliation{
$^1$Department of Physics, Wesleyan University, Middletown, Connecticut 06459, USA\\
$^2$Physics Department, Faculty of Mathematics and Physics, University of Ljubljana, Ljubljana, Slovenia\\
$^3$Max-Planck-Institute for Dynamics and Self-Organization, 37073 G\"ottingen, Germany
}

\begin{abstract}
We study the Loschmidt echo $F(t)$ for a class of dynamical systems showing critical chaos. Using 
a kicked rotor with singular potential as a prototype model, we found that the classical echo shows 
a gap (initial drop) $1-F_g$ where $F_g$ scales as $F_g(\alpha, \epsilon, \eta )= f_{\rm cl}(\chi_{
\rm cl} \equiv\eta^{3-\alpha}/\epsilon)$; $\alpha$ is the order of singularity of the potential, 
the spread of the initial phase space density and $\epsilon$ is the perturbation strength. Instead, 
the quantum echo gap is insensitive to $\alpha$, described by a scaling law $F_g = f_q(\chi_q = \eta^2/
\epsilon)$ which can be captured by a Random Matrix Theory modeling of critical systems. We trace this 
quantum-classical discrepancy to strong diffraction effects that dominate the dynamics.
\end{abstract}

\maketitle


The study of systems with a phase transition was always a fruitful subject of study for many areas
of theoretical and experimental physics. Specifically in the field of disordered metals, the celebrated 
Anderson Metal-Insulator Transition (MIT) \cite{EM08} has been an exciting subject of research for 
more than fifty years. On the other hand, the field of quantum chaos brought up a connection 
between quantized chaotic systems and localization ideas emerging from solid-state physics \cite{FGP82}. 
It has been shown that quantum suppression of classical diffusion is a result of wave 
interference phenomena of similar nature as the ones responsible for Anderson localization in disordered 
metals. Quite recently the connection between the two fields was further strengthen with the observation 
that certain non-KAM dynamical systems exhibiting classically anomalous diffusion, can have statistical 
properties resembling the ones of disordered metals at MIT \cite{MFDQS96,GW05}. This phenomenon is referred 
to as {\it critical chaos}. Some of these properties include, quantum anomalous diffusion \cite{CD88}, 
multifractal wavefunctions \cite{M00}, and critical spectral statistics \cite{KM97}. Many of 
these intriguing statistical properties can be exactly derived using non-conventional ensembles of 
random matrices with variance decaying from the diagonal in a power-law fashion \cite{EM00}, which in 
turn model a large variety of physical systems \cite{everybody}.

Most of the above studies discuss the stationary properties of critical systems. On the other hand
this knowledge is often not sufficient for a complete description of the dynamics. This need led us
in recent years to focus on new measures that efficiently probe the complexity of quantum time evolution.
One such measure is the so-called Loschmidt Echo (LE) which probes the sensitivity of quantum dynamics 
to external perturbations (for recent reviews see \cite{GPSZ06}). The recent literature on the subject 
is quite vast and ranges in areas as diverse as atomic optics \cite{GCZ97,AKD03,KASDMRKGRM03}, microwaves 
\cite{SGSS05}, elastic waves \cite{LW03}, quantum information \cite{NC00}, and quantum chaos \cite{JP01,
JSB01,JAB04,CT02,PS02,BC02,PLU95,VH03}. Formally, the LE $F(t)$, is defined as:
\begin{equation}
\label{eq:FidDef}
F(t) = |\bra{\psi_0}\eexp{iH_0t}\eexp{-iH_{\epsilon}t}\ket{\psi_0}|^2 ;\quad \hbar=1
\end{equation}
where $H_{\rm \epsilon}=H_{\rm 0}+\epsilon W$ is a one-parameter family of hamiltonians, $H_0$ is the
unperturbed hamiltonian, $V$ represents a perturbation of strength $\epsilon$ and $\ket{\psi_0}$ is 
an initial state. 

For a {\it non-critical} quantum system with a chaotic classical counterpart, the decay of the LE 
depends on the strength of the perturbation parameter $\epsilon$. Three regimes have been identified: 
the standard perturbative, the Fermi Golden Rule, and the non-perturbative regime. The first two can 
be described by Linear Response Theory leading to a decay which depends on the perturbation strength 
$\epsilon$ as $F(t)\sim\eexp{ -(\epsilon t)^2}$ and $F(t)\sim \eexp{-\epsilon^2t}$, respectively 
\cite{PS02,JSB01}. In the non-perturbative regime, the LE initially follows a Lyapunov decay $F(t)
\sim \eexp{ -\lambda t}$, with a rate given by the Lyapunov exponent $\lambda$ of the 
underlying classical system \cite{JP01,JSB01}, whereas for longer times (beyond the so-called
Ehrenfest time) the LE decays in accordance with the classical autocorrelation function \cite{BC02}. This 
behavior matches the decay of the classical echo according to the correspondence principle.

In this Letter we make the first step in understanding the echo decay of dynamical systems with 
critical chaos. The focus of the presentation is on the non-perturbative regime where our study 
revealed a novel result (see Fig. 1): we have discovered the appearance of an {\it echo gap} 
(initial drop of LE) $1-F_g$ for initial states which are distributed in the parts of phase-space 
where the hamiltonian function exhibits singularities. We have found that at the shortest classical
time scale (mean free time between singular scattering events) $F_g$ scales as
\begin{equation}
\label{fg}
F_g(\alpha, \epsilon, \eta ) =\left\{ 
\begin{array}{lcr}
f_{\rm cl}(\chi_{\rm cl})&{\rm where} & \chi_{\rm cl}=C\cdot \frac{\eta^{3-\alpha}}{\epsilon}\\
f_q(\chi_q)&{\rm where}&\chi_q=C\cdot \frac{\eta^2}{\epsilon}
\end{array}
\right.
\end{equation}
where the sub-indices ${\rm q/cl}$ indicate the quantum/classical scaling function for $F_g$. In Eq.
(\ref{fg}), $\alpha$ is the order of singularity of a non-analytical potential, $\eta$ is the 
characteristic spread of the phase space density of the initial classical or quantum state (e.g. its 
Wigner function) and the constant $C$ only depends on details of $H_0$ \cite{note1}. Moreover, we found 
that the scaling function $f(\chi)$ behaves asymptotically as $f(\chi \rightarrow 0)\sim \chi$. The above 
scaling laws were derived based on the analysis of classical dynamics and confirmed nicely by numerical 
simulations.

The apparent deviation of the quantum echo behavior from its classical counterpart (see Eq.(
\ref{fg})), can be seen as a violation of the quantum-classical correspondence; the latter being
confirmed in all previous fidelity studies. We have found that the origin of this anomalous 
behavior is due to strong diffraction effects which dictate the wave dynamics for the class of
dynamical systems we investigate in this Letter.

\begin{figure}
\includegraphics[width=1\columnwidth,keepaspectratio,clip]{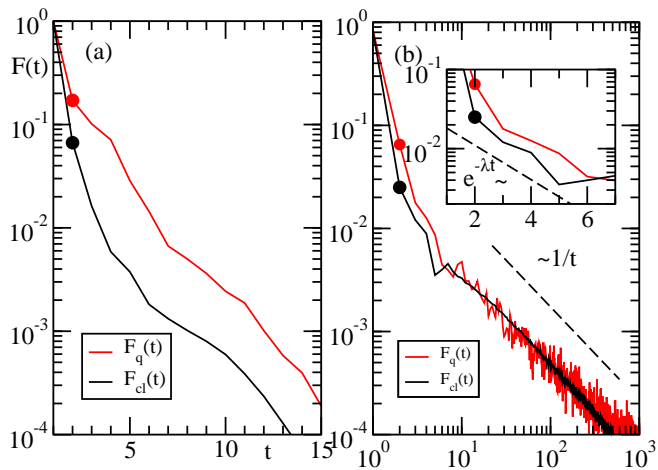}
\caption{\label{fig:fidelity}
Quantum (red line) and classical (black line) LE $F(t)$ for the KR with $V(q)=\log|q|$ and $K_0=1$ in 
the non-perturbative regime: (a) Torus geometry with $L=1$, $N=2^{17}$ and classical perturbation 
$\epsilon=10^{-4}$ (corresponding to $\sigma \approx 2.0$). The width of the initial preparation is 
$\eta \approx 0.04275$. (b) Cylinder geometry, with $L=10^3$, $N=2^{16}$ and classical perturbation 
$\epsilon=0.4$ (corresponding to $\sigma\approx 4.17$). One observes that after the initial Lyapunov 
decay (inset; the dashed line indicates an average Lyapunov decay), $F(t)$ follows a power law decay 
given by the autocorrelation function \cite{GW05,BC02}. Here we have $\eta \approx 0.15$. In both 
cases we have used more than $10^7$ trajectories for the classical calculation, while an averaging over 
$800$ initial Gaussian wavepackets centered at different momenta $p_0$ and $q_0=0$ has been done in the
quantum calculation. The filled circles indicate the first (classical or quantum) nontrivial timestep 
of $F_g=F(n=2)$. The statistical errors are smaller than the symbol size.
}
\end{figure}

\begin{figure}
\includegraphics[width=1\columnwidth,keepaspectratio,clip]{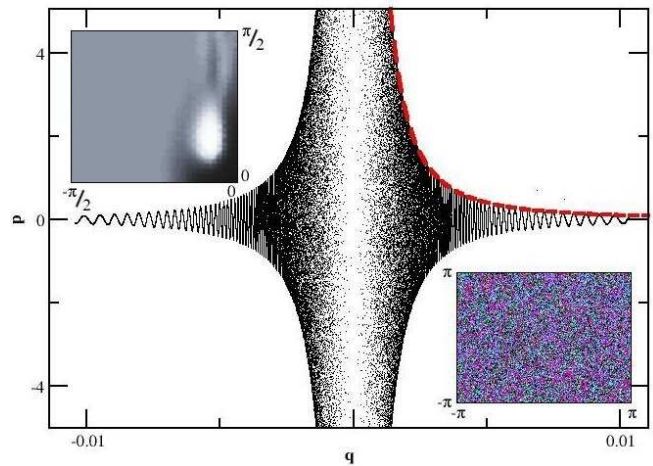}
\caption{\label{echomap}
Classical Echo map Eq. (\ref{echomap}) for the $\alpha=0, K_0=1$ singular potential (cylinder geometry
with $L=10^3$). The initial preparation is a box of size $\eta=\pi/300$ arround $(0,0)$. The evolution 
snapshot is for the shortest nontrivial time scale $n=2$. The red line is the theoretical
prediction of Eq.~(\ref{shift}). Upper left inset: The Wigner function representation (torus geometry) of 
the quantum echo map. The light areas correspond to negative phase space densities indicating diffraction 
phenomena. Lower right inset: A typical phase space for $K_0=1$ ($20$ trajectories involved
up to time $10^5$ iterations of the map)
}
\end{figure}

Below we consider a class of parametric Kicked Rotors (KR) defined by the time-dependent Hamiltonian 
\cite{GW05}
\begin{equation}
H_0=\frac{p^2}{2}+K_0 V(q)\sum_{n} \delta (t-nT)
\label{KRham}
\end{equation}
where $(p,q)$ is a pair of canonical variables, $T$ and $K_0$ are the period and the strength of the
kicking potential respectively. The class of KRs that we will study below have a potential which is 
given by 
\begin{equation}
\label{potential}
V(q) =\left\{ 
\begin{array}{lcr}
\abs{q}^{\alpha} & {\rm for} & \alpha\ne 0,\\
\log |q| & {\rm for} & \alpha= 0.\\
\end{array}
\right.
\end{equation}
The $\alpha=0$ case corresponds to MIT \cite{GW05}, and will be investigated below in detail. The 
classical dynamics is described by the following map:
\begin{equation}
\label{eq:map}
p_{n+1}= p_n - K_0 V'(q_n);\quad q_{n+1}= q_{n} + Tp_{n+1},
\end{equation}
where all variables are calculated immediately after one map iteration and $V'\equiv {\partial 
V(q)\over \partial q}$. 
The domain of $q$ is within the interval $-\pi<q<\pi$. The map
(\ref{eq:map}) can be studied on a cylinder $p\in \left(-\infty,\infty\right)$, which can also be 
closed to form a torus of length $2\pi L$, where $L$ is an integer. For $K_0>0$ the motion is chaotic
(see inset of Fig. 2) with a (local) Lyapunov exponent given by $\lambda (q) = 2\log(1+K_0/(2q^2) + 
\sqrt{K_0/q^2+[K_0/(2q^2)]^2})$ for $q\neq 0$. 

The quantum evolution is described by a a one-step unitary operator ${\hat U}_0$ acting on the 
wavefunction $\psi(q)$:
\begin{equation}
\label{Uop}
U_0=\exp(-i\tau {\hat n}^2/2)\, \exp(-i k V(q))\,,\quad \hbar = 1
\end{equation}
where ${\hat n}=-i\partial /\partial q$, $-N/2\leq n\leq N/2$, $\tau=(2\pi L/N)T$ and $k=(N/2\pi L)K_0$.
Optionally, we define an effective Planck constant $\hbar_{\rm eff} = 2\pi L/N$. The classical limit 
corresponds to $N\rightarrow \infty$. Without loss of generality we will assume below that $T=1$. It 
was shown in \cite{GW05}, that for $-0.5\leq \alpha\leq 0.5$ the eigenfunctions of the above unitary 
operator are multifractal while the levels resemble the statistical properties of disordered systems 
at MIT.

\begin{figure}
\includegraphics[width=1\columnwidth,keepaspectratio,clip]{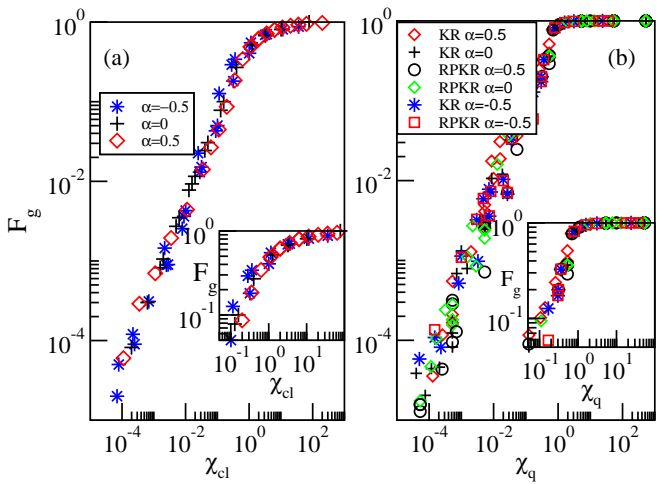}
\caption{\label{scaling}
The $F_g$ for various $\alpha,\eta, \epsilon$'s of the KR defined by Eqs. (\ref{KRham},\ref{potential}). 
In (a) we report the classical $F_g$ against the scaling variable $\chi_{\rm cl}$ (see Eq.~(\ref{fg})) 
while in (b) we report the corresponding quantum $F_g$, vs. the scaling variable $\chi_{\rm q}$ (see 
Eq.~(\ref{fg})). At the same sub-figure we report the results for the RPKR resulting from the model 
Eq. (\ref{Uop}) by a randomization of the phases of the
kinetic part of $U_0$. In the insets we report a magnification of the main panels in the regime of 
$F_g\approx 1$. In all cases an excellent data collapse is observed.
}
\end{figure}

For the echo calculation, we perturbed our system with the following (smooth) potential $W(q)= \cos{q}$. 
Correspondingly the perturbed quantum kicking parameter is $\sigma 
= \epsilon/\hbar_{\rm eff}$. Quantum mechanically, 
the initial preparation is a Gaussian wavepacket centered along the line of singularity i.e. $(q_0,p_0) 
= (0,p_0)$. The linear width $2\eta$ of the packet is taken to be minimal (i.e. $2\eta=\Delta p= \Delta q 
= \sqrt{\hbar_{\rm eff}/2}$), where we perform an averaging over different $p_0$'s in order 
to eliminate fluctuations. The corresponding classical initial preparation is given by a uniform distribution 
of trajectories located inside a box of area $A=2\eta\times 2\eta \sim \hbar_{\rm eff}$. We then define
the classical LE,  $F_{\rm cl}(t=n)$, as the overlap of the initial area $A_0$ with the area $\tilde{A}_f$ obtained 
by evolving $A_0$ for $n$ iterations of the perturbed map and then reversing the evolution for $n$ iterations 
with the unperturbed one. We have also checked that the results remain qualitatively the same when we chose 
an initial classical distribution to be a Gaussian density, equivalent to a Wigner function of the quantum 
Gaussian wavepacket.

We start our analysis with the classical derivation of Eq. (\ref{fg}). To this end we consider the 
classical echo dynamics \cite{VP04}. We denote by $\Phi_0(p,q)$ the forward symplectic map defined 
in Eq. (\ref{eq:map}) while by $\Phi_{\epsilon}=\Phi_0\circ P_{\epsilon}$ we denote the corresponding 
perturbed forward map. In this notation $P_{\epsilon}(p,q)= \left( p+\epsilon \sin(q),q\right)$ is 
a symplectic map generated by the perturbation (perturbation map). In this framework, the $n-$step 
echo map is defined as
\begin{equation}
\Phi_{n}^E\equiv \Phi_0^{-n}\circ \Phi_\epsilon^n =
 \tilde{P}_\epsilon^{(n-1)}\circ \tilde{P}_\epsilon^{(n-1)}\circ \cdots \circ \tilde{P}_\epsilon^{(0)}
\end{equation}
where the perturbation map is written in the interaction picture i.e. $\tilde{P}_\epsilon^{(n)} \equiv 
\Phi_0^{-n} \circ P_\epsilon \circ \Phi_0^n$. Explicitly, $(p_{n+1}^E,q_{n+1}^E) = \tilde{P}_\epsilon^{
(n)}(p_n^E,q_n^E)$ where
\begin{eqnarray}
 p_{n+1}^E &=& p_n^E +\epsilon \sin[\Phi_0^n(p_n^E,q_n^E)]_{q} 
{\partial [\Phi_0^n(p_n^E,q_n^E)]_{q}\over \partial q_n^E}\nonumber\\
 q_{n+1}^E &=& q_n^E -\epsilon \sin[\Phi_0^n(p_n^E,q_n^E)]_{q} 
{\partial [\Phi_0^n(p_n^E,q_n^E)]_{q}\over \partial p_n^E}
\label{echomap}
\end{eqnarray}

For singular potentials and initial conditions close to singularity, we express the phase space shift 
produced by the echo map after the second iteration step: 
\begin{eqnarray}
 \Delta p_{2}^E &=& p_2^E-p_0^E \approx -\epsilon \sin(q_1) K_0 V''(q_0)\nonumber\\
 \Delta q_{2}^E &=& q_2^E-q_0^E \approx -\epsilon \sin(q_1)
\label{shift}
\end{eqnarray}
where $q_1$ is given by the map (\ref{eq:map}) with the initial condition $(p_0^E,q_0^E)=(p_0,q_0)$. 
Due to the fact that the initial conditions are populating a box centered around the singularity 
line, we can assume that $\sin(q_1)$ is a pseudo-random variable with density ${\cal P}(x\equiv
\sin(q))=(1/\pi) (1-x^2)^{-1/2}$. The accuracy of our assumption is tested in Fig. 2, where we 
compare the envelope of the shift in momentum as it is given by Eq. (\ref{shift}) with the exact
echo dynamics.

For a given $q_0$ (and assuming $\epsilon\ll \eta$), the probability to return to the initial phase-
space box is estimated as
\begin{equation}
\label{prob}
{\cal P}(q_0)={1\over \pi}{\eta\over \eta + \epsilon K_0 |V''(q_0)|}.
\end{equation}
provided that the typical echo shift $\Delta p_2$ is much larger than $\eta$. The echo probability is 
just the integral of ${\cal P}(q_0)$ over the initial interval $q_0 \in [-\eta,\eta]$:
\begin{equation}
F_{\rm cl}^{g}= {1\over 2\eta}\int_{-\eta}^{\eta} {\cal P}(q_0)dq_0\approx 
{1\over 2\pi \epsilon K_0}\int_{-\eta}^{\eta} {dq_0 \over |V''(q_0)|}.
\label{clasfid}
\end{equation}

For the specific family of singular potentials discussed in this Letter, the above relation gives us:
\begin{equation}
\label{singfidcl}
F_{\rm cl}^g=\left\{ 
\begin{array}{lcr}
{1\over 2\pi K_0}{\eta^3\over \epsilon}&{\rm for}& \alpha=0\\
{1\over 2\pi K_0 |\alpha (\alpha-1)| (3-\alpha)} {\eta^{3-\alpha}\over \epsilon}&{\rm for}& \alpha\ne 0
\end{array}
\right.
\end{equation}

Our results in Eq. (\ref{singfidcl}) are nicely confirmed in Fig. 3, where we are plotting the echo gap for
various $\alpha,\epsilon$ and $\eta$ values by making use of the rescaled variable $\chi_{\rm cl}$ given 
by Eq. (\ref{fg}). Strictly speaking, the above results are applicable only for the case where $F_{\rm 
cl}^g\ll 1$. Nevertheless, our numerics indicates that the scaling behavior Eq. (\ref{fg}) continues 
to apply for values of $F_{\rm cl}^g \sim 1$.

We have also tested the results of the classical analysis against the quantum echo gap. A complete breakdown 
of the quantum-classical correspondence (QCC) is observed after the shortest non-trivial time scale (two 
iteration steps). This is associated with the fact that the Ehrenfest time for our system $t_E\sim \log(
h_{\rm eff})/\lambda(q_0)\rightarrow 0$ when $q_0\rightarrow 0$ \cite{BZ78}. Weak correspondence is restored 
for longer times when the echo dynamics spreads ergodically resulting in a vanishing measure of the critical 
line at $q_0=0$.

In Fig. 3 we observe that although the $\epsilon$ dependence of the quantum $F_g$ is captured by the classical 
calculations, both the $\eta\sim \sqrt{\hbar}$ and the $\alpha$ dependence differ drastically. The latter can 
be explained by a Random Phase Kicked Rotor (RPKR) with singular potential,
which is simply given by Eq. (\ref{Uop}) 
and replacing the eigenvalues of $\tau\hat{n}^2/2$ by random phases. This indicates that the appearance of a 
gap is insensitive to classical dynamics and thus can be captured by a Random Matrix Theory (RMT) modeling which 
preserves the power-law band structure of the evolution operator. Thus the fidelity gap is a universal phenomenon 
of critical systems described by these RMT models, and can be used as an alternative criterium to level or 
wavefunction statistics \cite{EM08,GW05,CD88,M00,KM97}.

We argue that the violation of the QCC is due to dominating diffraction effects that appear as a consequence
of the singular potential. This is illustrated in the inset of Figure 2 where we show the Wigner function
(computed according to Ref.~\cite{saraceno}) of the echo map for the first non-trivial time step. One 
observes the appearance of non-classical regions in the phase space where the Wigner function takes negative 
values.

In conclusion, we find that the LE of dynamical systems exhibiting critical chaos decays instantaneously with 
a gap that scales inverse proportionally to the strength of the perturbation. The order of the potential 
singularity is encoded in the scaling properties of the classical 
echo gap, while the corresponding quantum gap is insensitive to it and its scaling properties are described
by an RMT modeling of critical systems. This deviation is explained in the basis of strong diffraction which 
is a dominating mechanism of short time echo dynamics.

We thank G. S. Ng for useful discussions. The research was supported by the DFG FOR760 and by a grant from 
the United States-Israel Binational Science Foundation (BSF), Jerusalem, Israel.



\begin{thebibliography}{99}
\bibitem{note1} For the classical LE the dependence of $C$ on the order of the singularity $\alpha$ is
given in Eq.~(\ref{singfidcl}), while for the quantum LE we have found that $C$ is $\alpha$-independent.
\end{thebibliography}
\end{document}